\documentclass[twocolumn,10pt,aps,prd,preprintnumbers,showpacs,superscriptaddress,nofootinbib,amsmath,amssymb,floats,floatfix,showkeys,notitlepage,longbibliography]{revtex4-2}
\usepackage{orcidlink}
\usepackage{comment}
\usepackage{lipsum}
\usepackage{graphicx}
\usepackage{subfigure}
\usepackage{palatino}
\usepackage{sans}
\usepackage{hyperref}
\hypersetup{colorlinks=true,linkcolor=blue,urlcolor=blue,citecolor=blue}
\usepackage[toc,page]{appendix}
\usepackage[normalem]{ulem}
\usepackage{adjustbox}
\usepackage{latexsym}
\usepackage{amsmath}
\usepackage{amssymb}
\usepackage{amsfonts}
\usepackage{dcolumn}
\usepackage{bm}
\usepackage{tikz}
\usepackage{bigints}
\usepackage{array,tabularx,multirow,booktabs}
\usepackage[tracking=true]{microtype}
\usepackage{soul} 
\SetTracking{}{500}
\SetTracking{encoding={*}, shape=sc}{40}
\UseRawInputEncoding 
\allowdisplaybreaks
\usepackage{microtype}

\begin{document} \sloppy
\title{Phenomenology of Rotating GEUP Black Holes}

\author{Nikko John Leo S. Lobos \orcidlink{0000-0001-6976-8462}}
\email{nslobos@ust.edu.ph}
\affiliation{Electronics Engineering Department, University of Santo Tomas, Espa\~na Boulevard, Sampaloc, Manila 1008, Philippines}


\begin{abstract}
We investigate the phenomenological implications of quantum gravity on rotating black holes within the framework of the Generalized Extended Uncertainty Principle (GEUP), which incorporates both minimal length (ultraviolet) and large-scale (infrared) corrections. Lacking a full non-perturbative formulation of quantum gravity, we adopt a metric-based approach. We construct a stationary, axisymmetric ansatz via the Newman-Janis algorithm to model the kinematic features of a rotating black hole subject to Generalized Extended Uncertainty Principle (GEUP) corrections. The thermodynamic analysis reveals that in the infrared-dominated regime, the Hawking temperature scales as $T_H \sim M^{-3}$, leading to a rapid cooling phase that significantly prolongs the lifetime of supermassive black holes. We derive the modified Teukolsky Master Equation for gravitational perturbations and demonstrate that the background geometry preserves the isospectrality between axial and polar modes. In the eikonal limit, the quasinormal mode (QNM) spectrum exhibits orthogonal shifts: the minimal length parameter $\beta$ induces a spectral blueshift and enhanced damping, while the large-scale parameter $\alpha$ induces a spectral redshift and suppressed damping. We discuss how the thermodynamic evolution of the black hole breaks the geometric degeneracy that exists for static observers.
\end{abstract}


\maketitle

\section{Introduction}
\label{sec:intro}

The reconciliation of General Relativity (GR) with Quantum Mechanics remains one of the most profound challenges in theoretical physics. While GR successfully describes gravity at macroscopic scales, predicting the existence of black holes and gravitational waves \cite{LIGOScientific:2016aoc}, it inherently predicts its own breakdown at singularities where curvature divergences occur. Conversely, Quantum Field Theory (QFT) excels at microscopic scales but fails to incorporate gravity in a renormalizable manner. This tension suggests that the standard Heisenberg Uncertainty Principle (HUP) must be modified in the regime where gravitational effects become comparable to quantum fluctuations.

Various candidates for a theory of Quantum Gravity, such as String Theory, Loop Quantum Gravity (LQG), and Non-Commutative Geometry, seemingly converge on a fundamental prediction: the existence of a minimal measurable length, typically of the order of the Planck length $\ell_P \sim 10^{-35}$ m \cite{Gross:1987ar, Amati:1988tn, Garay:1994en}. This notion leads to the Generalized Uncertainty Principle (GUP), which modifies the commutation relations between position and momentum to include a minimal length scale \cite{Kempf:1994su, Maggiore:1993kv}. The phenomenological implications of GUP have been extensively studied in the context of black hole thermodynamics, leading to corrections in the Hawking temperature and potentially resolving the information loss paradox by preventing total evaporation \cite{Adler:2001vs, Scardigli:1999jh}.

However, the minimal length is not the only scale modification required. In contexts involving large length scales, such as physics in Anti-de Sitter (AdS) space or expanding cosmological backgrounds, the uncertainty principle may also require an infrared (IR) correction. This leads to the Extended Uncertainty Principle (EUP), which introduces a minimum momentum uncertainty \cite{Bambi:2007ty, Park:2007az}. When combined, these UV and IR modifications form the Generalized Extended Uncertainty Principle (GEUP). The GEUP framework is particularly compelling because it captures the "UV/IR mixing" phenomena characteristic of non-local field theories and provides a more complete deformation of the phase space structure \cite{Mureika:2018gxl}.

Recent studies have addressed this by constructing effective metrics with quantum corrections \cite{Xiang:2009yq,Scardigli:2014qka}. However, a comprehensive analysis of the GEUP effects on the stability and dynamic ringdown of rotating black holes remains scarce.

In this work, we address this gap by constructing the metric for a rotating black hole explicitly deformed by the GEUP. We investigate the physical consequences of this deformation through two complementary windows: thermodynamics and gravitational wave spectroscopy. First, we analyze the thermal stability of the black hole, demonstrating how the interplay between the GUP parameter ($\beta$) and the EUP parameter ($\alpha$) shifts the phase transition critical points. Second, utilizing the Eikonal limit, we calculate the Quasinormal Mode (QNM) frequencies. We show that $\alpha$ and $\beta$ induce opposing shifts in the complex frequency plane—leading to a spectral redshift and blueshift, respectively. This orthogonality offers a theoretical pathway to disentangle minimal length (UV) effects from large-scale (IR) modifications using future high-precision ringdown data.

Our approach aligns with the parameterized deviation frameworks often employed in gravitational wave and shadow analysis. Rather than deriving a solution from a specific modified Lagrangian, which remains ambiguous in GEUP frameworks, we propose a physically motivated geometry. This allows us to map the GEUP parameters ($\alpha, \beta$) directly to observables, providing a testbed for constraining minimal length theories with astrophysical data.

The paper is organized as follows: Section \ref{sec:metric} establishes the modified commutation relations and derives the effective GEUP mass and metric. Section \ref{sec:thermo} analyzes the thermodynamic quantities, presenting the heat capacity and stability profiles. Section \ref{sec:perturbations} investigates the gravitational wave signatures via the QNM spectrum. Section \ref{sec:shadow} investigates the observational consistency of the model and discusses how the thermodynamic evolution breaks the geometric degeneracy. Finally, Section \ref{conc} summarizes our findings and discusses future outlooks.

\section{The Metric: Rotating GEUP Black Hole}
\label{sec:metric}

In this section, we construct the spacetime metric for a rotating black hole satisfying the Generalized Extended Uncertainty Principle (GEUP). We adopt the mass renormalization formalism proposed by Mureika \cite{Mureika:2019}, which accounts for large-scale quantum gravity corrections while preserving the asymptotic flatness of the spacetime. The stationary, axisymmetric solution is subsequently generated via the Newman-Janis Algorithm (NJA) \cite{Newman:1965tw}.

The Generalized Extended Uncertainty Principle modifies the standard Heisenberg algebra by introducing characteristic length scales at both the ultraviolet (Planck scale) and infrared (cosmological scale) limits. The position-momentum uncertainty relation is expressed as \cite{Mureika:2019}:
\begin{equation}
\Delta x \Delta p \ge \frac{\hbar}{2} \left( 1 + \beta l_{Pl}^2 \frac{(\Delta p)^2}{\hbar^2} + \alpha \frac{(\Delta x)^2}{L_*^2} \right),
\label{eq:GEUP_uncertainty}
\end{equation}
where $l_{Pl}$ is the Planck length, $L_*$ is a fundamental large-scale length parameter, and $\alpha, \beta$ are dimensionless deformation constants.

Within the corpuscular gravity framework, the black hole is treated as a condensate of gravitons. Mureika \cite{Mureika:2019} demonstrated that imposing the uncertainty relation (Eq. \ref{eq:GEUP_uncertainty}) on the Compton-Schwarzschild correspondence modifies the relationship between the horizon radius and the geometric mass. This modification is physically realized as a renormalization of the mass parameter $M$ into an effective mass $\mathcal{M}$. Incorporating both the GUP (small-scale) and EUP (large-scale) corrections, the effective mass is given by:
\begin{equation}
\mathcal{M} = M \left( 1 + \frac{\beta_{0} M_{Pl}^2}{2 M^2} + \frac{\alpha_{0} G^2 M^2}{L_*^2} \right),
\label{eq:effective_mass}
\end{equation}
where $\beta_0$ and $\alpha_0$ represent the rescaled coupling constants. The term proportional to $M^{-2}$ arises from GUP corrections dominating at the Planck scale, while the term proportional to $M^2/L_*^2$ arises from EUP corrections, which become significant for supermassive black holes where the gravitational radius approaches $L_*$.

The corresponding static, spherically symmetric vacuum solution in Schwarzschild coordinates $(t, r, \theta, \phi)$ is therefore:
\begin{equation}
ds^2_{\text{static}} = -f(r) dt^2 + \frac{1}{f(r)} dr^2 + r^2 d\Omega^2,
\label{eq:static_metric}
\end{equation}
with the metric function:
\begin{equation}
f(r) = 1 - \frac{2G\mathcal{M}}{r}.
\label{eq:f_r_GEUP}
\end{equation}
In this framework, the quantum gravity corrections are naturally incorporated into the \textbf{constant effective mass} $\mathcal{M}$. This approach preserves the algebraic structure of the Schwarzschild solution while modifying the strength of the gravitational potential to account for both short-distance (GUP) and large-distance (EUP) deformations.

To introduce angular momentum $a$, we employ the Newman-Janis Algorithm. We first transform the static metric (Eq. \ref{eq:static_metric}) to advanced Eddington-Finkelstein coordinates $(u, r, \theta, \phi)$ via $du = dt - f(r)^{-1}dr$. The contravariant metric components are decomposed into a null tetrad basis $(l^\mu, n^\mu, m^\mu, \bar{m}^\mu)$:
\begin{equation}
\begin{aligned}
l^\mu &= \delta^\mu_r, \\
n^\mu &= \delta^\mu_u - \frac{1}{2} \left( 1 - \frac{2G\mathcal{M}}{r} \right) \delta^\mu_r, \\
m^\mu &= \frac{1}{\sqrt{2}r} \left( \delta^\mu_\theta + \frac{i}{\sin\theta} \delta^\mu_\phi \right).
\end{aligned}
\end{equation}
We perform the complex coordinate transformation $r \to r' = r + ia\cos\theta$ and $u \to u' = u - ia\cos\theta$. Since the effective mass $\mathcal{M}$ is a constant parameter dependent only on the source mass $M$ and fundamental constants, it remains invariant under the complexification of the radial coordinate. The transformation acts solely on the radial potential term $1/r$:
\begin{equation}
\frac{1}{r} \longrightarrow \frac{1}{2} \left( \frac{1}{r'} + \frac{1}{\bar{r}'} \right) = \frac{r}{\Sigma},
\end{equation}
where $\Sigma \equiv r^2 + a^2 \cos^2\theta$. The transformed null tetrad vectors are:
\begin{equation}
\begin{aligned}
l'^\mu &= \delta^\mu_r, \\
n'^\mu &= \delta^\mu_u - \frac{1}{2} \left( 1 - \frac{2G\mathcal{M} r}{\Sigma} \right) \delta^\mu_r, \\
m'^\mu &= \frac{1}{\sqrt{2}(r + ia\cos\theta)} \left( \delta^\mu_\theta + \frac{i}{\sin\theta} \delta^\mu_\phi \right).
\end{aligned}
\end{equation}
Reconstructing the metric tensor $g^{\mu\nu}$ and transforming to Boyer-Lindquist coordinates yields the line element for the rotating GEUP black hole:
\begin{equation}
\begin{aligned}
ds^2 = &-\left( 1 - \frac{2G\mathcal{M}r}{\Sigma} \right) dt^2 + \frac{\Sigma}{\Delta} dr^2 + \Sigma d\theta^2 \\
&+ \left( (r^2+a^2)^2 - \Delta a^2 \sin^2\theta \right) \frac{\sin^2\theta}{\Sigma} d\phi^2 \\
&- \frac{4G\mathcal{M} a r \sin^2\theta}{\Sigma} dt d\phi,
\end{aligned}
\label{eq:rotating_metric}
\end{equation}
where the horizon function $\Delta$ is defined as:
\begin{equation}
\Delta = r^2 - 2G\mathcal{M}r + a^2.
\label{eq:Delta_func}
\end{equation}
It is important to note that because the effective mass $\mathcal{M}$ is treated as a global constant determined by the source parameters $(M, \alpha, \beta)$ rather than a coordinate-dependent function, the metric in Eq. (\ref{eq:rotating_metric}) retains the vacuum character of the standard Kerr solution. Direct calculation of the Einstein tensor yields $G_{\mu\nu} = 0$, implying that the leading-order components of the effective energy-momentum tensor vanish ($\rho = p_i = 0$). Consequently, the Weak Energy Condition ($\rho \ge 0, \rho + p_i \ge 0$) is trivially satisfied throughout the spacetime, ensuring that the phenomenological modifications do not introduce unphysical matter distributions in the domain of outer communication.

The derived metric is formally identical to the Kerr solution, with the standard geometric mass replaced by the GEUP effective mass $\mathcal{M}$. We emphasize, however, that the Newman-Janis Algorithm is not guaranteed to yield exact solutions to modified field equations \cite{Hansen:2013owa}. Therefore, we treat Eq. (\ref{eq:rotating_metric}) not as a vacuum solution to a specific effective action, but as a phenomenological ansatz defined by its symmetries. This metric-based approach allows us to constrain the deformation parameters $\alpha$ and $\beta$ via kinematic observables (shadows, QNMs) irrespective of the underlying dynamics.

The event horizons are defined by the roots of $\Delta(r) = 0$, yielding:
\begin{equation}
r_{\pm} = G\mathcal{M} \pm \sqrt{G^2\mathcal{M}^2 - a^2}.
\label{eq:horizons}
\end{equation}
In the regime of supermassive black holes ($M \gg M_{Pl}$), the EUP correction dominates, and $\mathcal{M} \approx M(1 + \epsilon_{EUP})$. This results in an expansion of the event horizon radius $r_+$ relative to the General Relativistic prediction for a fixed bare mass $M$. Consequently, the Bekenstein-Hawking entropy, proportional to the horizon area $A_H = 4\pi (r_+^2 + a^2)$, is enhanced by large-scale uncertainty correlations.

The stationary limit surface, defining the outer boundary of the ergosphere, is located at:
\begin{equation}
r_{sl}(\theta) = G\mathcal{M} + \sqrt{G^2\mathcal{M}^2 - a^2 \cos^2\theta}.
\end{equation}
The volume of the ergosphere is sensitive to the mass parameter. Since $\mathcal{M} > M$ in the EUP regime, the GEUP corrections enlarge the ergosphere. This modification implies an increased capacity for energy extraction via the Penrose process. Furthermore, the extremal limit for the spin parameter is shifted to $a_{max} = G\mathcal{M}$. Since $\mathcal{M} > M$, a GEUP black hole can support a higher angular momentum than a standard Kerr black hole of identical baryonic mass without violating the cosmic censorship conjecture.

\section{Thermodynamic Properties and Cryogenic Evolution}
\label{sec:thermo}

In this section, we analyze the thermodynamic structure of the Rotating GEUP black hole. Since the GEUP metric derived in Section \ref{sec:metric} retains the algebraic form of the Kerr solution through the effective mass renormalization $M \to \mathcal{M}$, the standard geometric definitions of thermodynamic quantities remain applicable \cite{Mureika:2019}. However, the dependence of $\mathcal{M}$ on the fundamental uncertainty parameters introduces significant deviations in the thermal evolution of the black hole, particularly for supermassive systems where large-scale correlations become dominant.

\subsection{Thermodynamic Variables and the First Law}

Throughout this analysis, we identify $\mathcal{M}$ as the total ADM mass (internal energy) of the spacetime observed at infinity. The parameter $M$ denotes the bare source mass. Consequently, the First Law of Thermodynamics is satisfied by the physical variables: $d\mathcal{M} = T_H dS + \Omega_H dJ$. 

The Hawking temperature $T_H$ is determined by the surface gravity $\kappa$ at the outer event horizon $r_+$. For a stationary, axisymmetric spacetime, the surface gravity is defined via the Killing vector $\chi^\mu = \xi^\mu_{(t)} + \Omega_H \xi^\mu_{(\phi)}$, such that $\kappa^2 = -\frac{1}{2} (\nabla_\mu \chi_\nu)(\nabla^\mu \chi^\nu)|_{r_+}$. This yields the standard relation renormalized by $\mathcal{M}$:
\begin{equation}
T_H = \frac{\kappa}{2\pi} = \frac{r_+ - G\mathcal{M}}{4\pi (r_+^2 + a^2)},
\label{eq:temperature_def}
\end{equation}
where the horizon radius is $r_+ = G\mathcal{M} + \sqrt{G^2\mathcal{M}^2 - a^2}$. Substituting the effective mass $\mathcal{M}$ from Eq. (\ref{eq:effective_mass}), we can express the temperature in terms of the dimensionless spin parameter $\tilde{a} = a/G\mathcal{M}$:
\begin{equation}
T_H = \frac{1}{4\pi G \mathcal{M}} \left( \frac{\sqrt{1 - \tilde{a}^2}}{1 + \sqrt{1 - \tilde{a}^2}} \right).
\label{eq:temperature_explicit}
\end{equation}

Equation (\ref{eq:temperature_explicit}) reveals a crucial physical implication of the GEUP framework. In the EUP-dominated regime ($M \gg M_{Pl}$), the effective mass scales as $\mathcal{M} \sim M^3/L_*^2$ \cite{Mureika:2019}. Consequently, the temperature scales as $T_H \sim \mathcal{M}^{-1} \sim M^{-3}$, falling much more rapidly than the standard Schwarzschild scaling ($T_H \sim M^{-1}$). This implies that large-scale quantum correlations effectively "cool" supermassive black holes, significantly extending their evaporation lifetime compared to General Relativistic predictions.

The Bekenstein-Hawking entropy is proportional to the horizon area $A_H$. Unlike approaches that modify the area law itself, the Mureika formalism encapsulates the deformation within the geometry:
\begin{equation}
S = \frac{A_H}{4G} = \frac{4\pi (r_+^2 + a^2)}{4G} = \frac{2\pi G \mathcal{M}^2}{1} \left( 1 + \sqrt{1 - \tilde{a}^2} \right).
\label{eq:entropy}
\end{equation}
While the entropy follows the standard area law with respect to the effective mass, its relation to the bare mass $M$ is super-extensive. For large $M$, $S \propto \mathcal{M}^2 \propto M^6$. This drastic enhancement of the density of states indicates that the EUP modification allows the horizon to encode significantly more information than the standard holographic bound would suggest.

Finally, we identify the angular velocity of the horizon, $\Omega_H$, which represents the conjugate potential to the angular momentum $J = a\mathcal{M}/G$:
\begin{equation}
\Omega_H = \frac{a}{r_+^2 + a^2} = \frac{\tilde{a}}{2G\mathcal{M} (1 + \sqrt{1-\tilde{a}^2})}.
\end{equation}
We verify that these quantities satisfy the First Law of Thermodynamics for the effective system:
\begin{equation}
d\mathcal{M} = T_H dS + \Omega_H dJ.
\label{eq:first_law}
\end{equation}
This consistency confirms that the formalism preserves the Legendre transformation structure of black hole thermodynamics provided the variations are taken with respect to the renormalized mass parameter.

\subsection{Heuristic Derivation of Temperature Limits}
The Hawking temperature can also be understood heuristically via the uncertainty principle. In the GEUP framework, the uncertainty relation $\Delta x \Delta p \gtrsim \hbar (1 + \alpha (\Delta x)^2)$ implies a modification to the temperature-mass relation. Identifying the position uncertainty with the horizon radius $\Delta x \sim r_+ \approx 2M$, the modified temperature is:
\begin{equation}
T_{H} \approx \frac{1}{8\pi M} \left[ 1 + \alpha (2M)^2 \right]^{-1} \approx \frac{1}{8\pi M (1 + 4\alpha M^2)}.
\label{eq:temp_scaling_heuristic}
\end{equation}
This expression recovers the standard Hawking temperature $T_{GR} = (8\pi M)^{-1}$ in the limit $\alpha \to 0$. However, distinct behaviors emerge in the asymptotic regimes. For small masses (GUP regime), the temperature reaches a maximum, while for supermassive black holes (EUP regime, $M \gg 1/\sqrt{\alpha}$), the term in the denominator dominates, leading to the scaling:
\begin{equation}
T_{H} \xrightarrow{M \to \infty} \frac{1}{32\pi \alpha M^3}.
\label{eq:temp_limit}
\end{equation}
This $T \sim M^{-3}$ behavior indicates that large GEUP black holes are significantly colder than their General Relativistic counterparts.

\begin{figure}[htbp]
    \centering
    \includegraphics[width=0.48\textwidth]{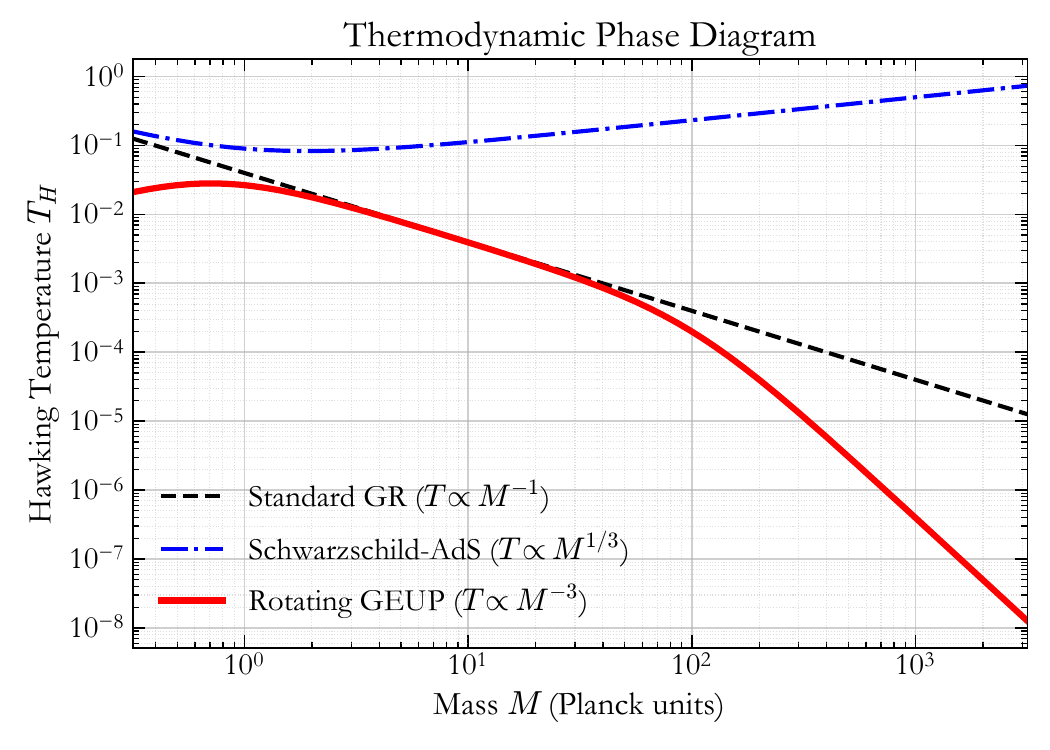}
    \caption{Log-log phase diagram comparing the Hawking temperature evolution for three black hole models. The dashed black line represents standard General Relativity ($T \propto M^{-1}$). The dot-dashed blue line represents the Schwarzschild-AdS behavior, where large masses become hot and thermodynamically stable ($T \propto M^{1/3}$). The solid red line depicts the Rotating GEUP black hole. Key features include the maximum temperature peak at the Planck scale (GUP regime) and the rapid cooling phase at large scales (EUP regime) where $T \propto M^{-3}$, contrasting sharply with the AdS behavior.}
    \label{fig:3}
\end{figure}

To visualize the distinct thermodynamic behavior of the GEUP black hole, we present the temperature-mass phase diagram in Figure \ref{fig:3}. The plot compares the GEUP trajectory against the standard Schwarzschild behavior~\cite{Hawking:1975vcx} and the Schwarzschild-AdS case~\cite{Hawking:1982dh}. In the intermediate mass regime, the curve follows the standard General Relativistic scaling ($T_H \propto M^{-1}$). However, significant deviations appear in the asymptotic limits. At the Planck scale (left), the GUP correction forces the temperature to reach a maximum and vanish as $M \to M_{min}$, preventing the final evaporation singularity. Conversely, in the infrared regime (right), the trajectory diverges sharply from both the standard case and the AdS case. While large AdS black holes become thermodynamically stable with positive specific heat ($T_{AdS} \propto M^{1/3}$), the GEUP black hole enters a 'super-cooling' phase where $T_{GEUP} \propto M^{-3}$. This steep negative slope visually confirms the rapid suppression of Hawking radiation for supermassive sources.

\subsection{Thermal Stability Analysis}

The thermodynamic stability of the black hole is characterized by its heat capacity. For a rotating system, we consider the heat capacity at constant angular momentum, defined as $C_J = T_H \left( \frac{\partial S}{\partial T_H} \right)_J$. Following the method of Davies \cite{Davies:1977}, the divergence of this quantity signals a phase transition. Using the chain rule with respect to the horizon radius $r_+$, we derive:
\begin{equation}
C_J = \frac{2\pi (r_+^2 + a^2)^2 (G\mathcal{M} r_+ - a^2)}{r_+^2 (3G\mathcal{M} - 2r_+) - a^2(G\mathcal{M} - 2r_+)}.
\label{eq:heat_capacity}
\end{equation}
The sign of $C_J$ determines stability: $C_J < 0$ indicates instability (typical of Schwarzschild black holes), while $C_J > 0$ indicates stability. The roots of the denominator in Eq. (\ref{eq:heat_capacity}) correspond to Davies points, marking phase transitions between stable and unstable configurations; the GEUP parameters $\alpha$ and $\beta$ shift these critical points, altering the domain of stability in the $(M, a)$ parameter space.

\begin{figure}[htbp]
    \centering
    \includegraphics[width=0.48\textwidth]{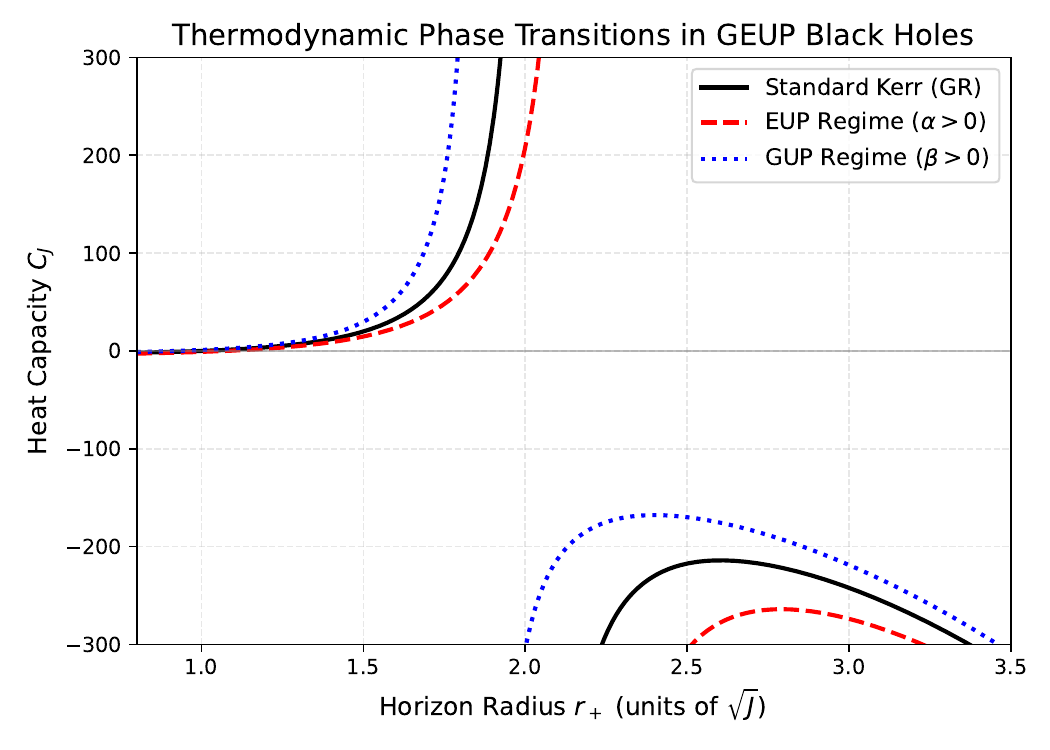}
    \caption{Behavior of the heat capacity $C_{J}$ (in units of $M_{Pl}^2$) as a function of the horizon radius $r_{+}$ (normalized by the bare mass $GM$). The vertical asymptotes correspond to Davies points, marking the phase transition. We adopt the standard thermodynamic sign convention where $C_{J} > 0$ indicates stability and $C_{J} < 0$ indicates instability. The GEUP corrections shift the location of these critical points, extending the domain of stability compared to the standard Kerr metric.}
    \label{fig:heat_capacity}
\end{figure}

\subsection{Cryogenic Evolution and Lifetime Enhancement}
To verify the physical implications of the modified temperature scaling, we examine the evaporation rate of the black hole. The black hole radiates energy at a rate governed by the Stefan-Boltzmann law. Since the effective mass $\mathcal{M}$ represents the total ADM mass (energy) of the spacetime, the energy loss equation is:
\begin{equation}
\frac{d\mathcal{M}}{dt} \approx - \sigma A_H T_H^4,
\label{eq:22}
\end{equation}
where $\sigma$ is the Stefan-Boltzmann constant and $A_H$ is the horizon area.

For a supermassive black hole in the EUP-dominated regime ($M \gg M_{Pl}$), we established in Eq. (\ref{eq:temp_limit}) that the temperature scales anomalously as $T_H \propto M^{-3}$, while the area scales approximately as $A_H \propto M^2$ (assuming $\mathcal{M} \approx M$ for the geometric cross-section to first order). Substituting these scalings into the luminosity equation:
\begin{equation}
\mathcal{L} = -\frac{d\mathcal{M}}{dt} \propto M^2 (M^{-3})^4 \propto M^{-10}.
\end{equation}
This stands in stark contrast to standard General Relativity, where $\mathcal{L}_{GR} \propto M^{-2}$. The GEUP black hole is significantly dimmer and colder than its GR counterpart.

The evaporation lifetime $\tau$ is obtained by integrating the mass loss:
\begin{equation}
\tau \approx - \int_{M_i}^{0} \frac{1}{\mathcal{L}} \frac{d\mathcal{M}}{dM} dM \approx \int_{0}^{M_i} M^{10} dM \propto M_i^{11}.
\end{equation}
The EUP modification acts as a `cryogenic' mechanism, suppressing the evaporation rate for large black holes. We note that this behavior contrasts with black holes in Anti-de Sitter (AdS) space, where the large-scale potential typically induces positive specific heat ($T \sim M^{1/3}$). Here, the EUP-induced renormalization preserves the negative specific heat of the system but drastically enhances the cooling rate ($T \sim M^{-3}$), thereby prolonging the semi-classical lifetime.

Physically, this rapid cooling behavior ($T_H \sim M^{-3}$) can be understood as a consequence of the macroscopic `fuzziness' introduced by the Extended Uncertainty Principle in the infrared regime. Unlike standard General Relativity, where the horizon radius is strictly proportional to the bare mass ($r_+ \propto M$), the GEUP correction implies that position uncertainty grows with the scale of the system. For supermassive black holes, this macroscopic uncertainty effectively delocalizes the horizon, causing the effective gravitational radius to inflate disproportionately ($\mathcal{M} \sim M^3$). This geometric inflation drastically reduces the surface gravity ($\kappa \propto 1/r_+$), driving the black hole into a `cryogenic' state much faster than the standard Hawking scaling would predict.

\section{Gravitational Perturbations}
\label{sec:perturbations}

Having established the thermodynamic stability of the Rotating GEUP black hole, we now turn to its dynamical stability and gravitational wave signature by studying the propagation of gravitational perturbations. Since the GEUP metric (Eq. \ref{eq:rotating_metric}) is isometric to the Kerr metric under the mapping $M \to \mathcal{M}$. Spacetime remains algebraically special (Petrov Type D). Crucially, this algebraic classification relies on our treatment of the renormalized mass $\mathcal{M}$ as a global constant fixed by the source’s bare mass, rather than a running coupling function of the radial coordinate $\mathcal{M}(r)$. This assumption preserves the Type D character of the background geometry, thereby allowing us to employ the standard Newman-Penrose formalism \cite{Newman:1961} to decouple the perturbation equations—a procedure first developed by Teukolsky \cite{Teukolsky:1973}.

\subsection{The Teukolsky Master Equation}
We consider perturbations of the Weyl scalar $\Psi_4$, which describes outgoing gravitational radiation in the wave zone. In the Newman-Penrose formalism, $\Psi_4$ is defined as the contraction of the Weyl tensor $C_{\alpha\beta\gamma\delta}$ with the complex null vector $\bar{m}^\mu$:
\begin{equation}
\Psi_4 = -C_{\alpha\beta\gamma\delta} n^\alpha \bar{m}^\beta n^\gamma \bar{m}^\delta.
\end{equation}
For a vacuum Type D background, the linearized equation of motion for a field of spin-weight $s$ (where $s=-2$ for gravitational perturbations) separates into the Teukolsky Master Equation \cite{Teukolsky:1973}:
\begin{equation}
\begin{aligned}
&\left[ \frac{\partial}{\partial r} \Delta \frac{\partial}{\partial r} - \frac{1}{\Delta} \left\{ (r^2+a^2) \frac{\partial}{\partial t} + a \frac{\partial}{\partial \phi} - s\Delta'(r) \right\}^2 \right. \\
&+ \left. 4s(r+ia\cos\theta) \frac{\partial}{\partial t} + \frac{1}{\sin\theta} \frac{\partial}{\partial \theta} \sin\theta \frac{\partial}{\partial \theta} \right. \\
&+ \left. \left( a\sin\theta \frac{\partial}{\partial t} + \frac{1}{\sin\theta} \frac{\partial}{\partial \phi} - is\cos\theta \right)^2 \right] \Psi = 0,
\end{aligned}
\label{eq:teukolsky_master}
\end{equation}
where $\Psi = \rho^{-4} \Psi_4$ is the master field variable, and $\rho = -1/(r-ia\cos\theta)$.

The uncertainty principle corrections enter this equation explicitly through the horizon function $\Delta(r)$. Recall from Section \ref{sec:metric} that for the GEUP black hole:
\begin{equation}
\Delta(r) = r^2 - 2G\mathcal{M}r + a^2,
\label{eq:delta_geup}
\end{equation}
where $\mathcal{M} = M(1 + \beta_0 M_{Pl}^2/2M^2 + \alpha_0 G^2 M^2/L_*^2)$ is the effective mass. Unlike other modified gravity theories that introduce higher-order derivative terms or break Lorentz invariance, the Mureika formalism preserves the second-order hyperbolic nature of the wave equation. This guarantees the well-posedness of the Cauchy problem for gravitational radiation in the GEUP framework.

\subsection{Separation of Variables}

We seek separable solutions of the form:
\begin{equation}
\Psi(t, r, \theta, \phi) = e^{-i\omega t} e^{im\phi} R_{\ell m}(r) S_{\ell m}(\theta),
\end{equation}
where $\omega$ is the (complex) quasinormal frequency, $m$ is the azimuthal quantum number, and $\ell$ is the multipole index ($\ell \ge 2$). Substituting this ansatz into Eq. (\ref{eq:teukolsky_master}) decouples the system into angular and radial ordinary differential equations.

The angular function $S_{\ell m}(\theta)$ satisfies the Spin-Weighted Spheroidal Harmonic equation:
\begin{equation}
\begin{split}
&\frac{1}{\sin\theta} \frac{d}{d\theta} \left( \sin\theta \frac{dS_{\ell m}}{d\theta} \right) \\&+
 a^2\omega^2 \cos^2(\theta)S_{\ell m} - \frac{m^2}{\sin^2\theta}S_{\ell m} \\&- 2a\omega s \cos(\theta)S_{\ell m} + 4a\omega \cot(\theta )sS_{\ell m} \\&+ sS_{\ell m} + A_{\ell m}S_{\ell m}  = 0,
\end{split}
\label{eq:angular_ode}
\end{equation}
where $A_{\ell m}$ is the separation constant. It is important to note that the angular equation depends on the rotation parameter $a$ and the frequency $\omega$, but \textit{not} explicitly on the mass $\mathcal{M}$. Therefore, the angular eigenfunctions (spheroidal harmonics) for the GEUP black hole are identical to those of the Kerr black hole. The uncertainty principle affects the angular sector only implicitly through the shift in the resonant frequencies $\omega$.

The radial function $R_{\ell m}(r)$ satisfies the modified Teukolsky Radial Equation:
\begin{equation}
\begin{split}
&\Delta^{-s} \frac{d}{dr} \left( \Delta^{s+1} \frac{dR_{\ell m}}{dr} \right) \\&+ \left( \frac{K^2 - 2is(r-G\mathcal{M})K}{\Delta} + 4is\omega r - \lambda \right) R_{\ell m} = 0,
\end{split}
\label{eq:radial_ode}
\end{equation}
where $K \equiv (r^2+a^2)\omega - am$, and $\lambda \equiv A_{\ell m} + a^2\omega^2 - 2am\omega$.
This equation governs the generation and propagation of gravitational waves. The GEUP corrections are encoded in two terms:
1.  The function $\Delta(r)$, which determines the location of the horizon and the potential barrier.
2.  The explicit appearance of $\mathcal{M}$ in the potential term $2is(r-G\mathcal{M})K$. 

\subsection{Physical Implications for Gravitational Waves}

The mathematical structure of the Radial Equation (\ref{eq:radial_ode}) permits a rigorous derivation of the spectral shifts and symmetry properties induced by the GEUP corrections. We analyze two distinct signatures: the shifting of the Quasinormal Mode (QNM) frequencies and the preservation of isospectrality.

\subsubsection{QNM Frequency Shift via the Eikonal Limit}

To quantify the shift in the gravitational wave spectrum, we consider the Eikonal (geometric optics) limit ($\ell \gg 1$). In this regime, the Quasinormal Modes are intimately related to the properties of the unstable photon orbit (light ring) encircling the black hole. As established by Ferrari and Mashhoon \cite{Ferrari:1984zz}, and later generalized by Cardoso et al. \cite{Cardoso:2008bp}, the complex QNM frequency $\omega = \omega_R + i\omega_I$ corresponds to the orbital frequency and the instability timescale of null geodesics:
\begin{equation}
\omega_{n\ell} \approx \Omega_c \ell - i \left(n + \frac{1}{2}\right) |\lambda_L|,
\label{eq:eikonal_relation}
\end{equation}
where $\Omega_c$ is the angular velocity of the photon sphere and $\lambda_L$ is the Lyapunov exponent characterizing the instability of the orbit. 

We begin with the geodesic Hamiltonian for null particles ($H = \frac{1}{2}g^{\mu\nu}p_\mu p_\nu = 0$). For equatorial orbits ($\theta = \pi/2$), the radial motion is governed by an effective potential $V_{geo}(r)$:
\begin{equation}
\dot{r}^2 = V_{geo}(r) = E^2 - V_{eff}(r)
\end{equation}
with
\begin{equation}  
\quad V_{eff}(r) = -\frac{g_{tt} L^2 + 2g_{t\phi} E L + g_{\phi\phi} E^2}{g_{rr} g_{\phi\phi} - g_{t\phi}^2} g_{rr}.
\end{equation}
Introducing the impact parameter $b \equiv L/E$, the condition for an unstable circular orbit requires the potential and its derivative to vanish simultaneously:
\begin{equation}
V_{geo}(r_c) = 0 \quad \text{and} \quad \frac{dV_{geo}}{dr}\bigg|_{r_c} = 0.
\label{eq:photon_sphere_cond}
\end{equation}
For our Rotating GEUP metric, substituting the metric components derived in Section \ref{sec:metric}, Eq. (\ref{eq:photon_sphere_cond}) yields a generalized polynomial equation for the photon sphere radius $r_c$. To leading order in the rotation $a$ and GEUP corrections, this radius is shifted as:
\begin{equation}
r_c \approx 3M \left( 1 - \frac{2}{3\sqrt{3}}\frac{a}{M} - \frac{\beta}{18} + \frac{27}{2}\alpha M^2 \right).
\end{equation}
The real part of the frequency, determined by the orbital angular velocity $\Omega_c = \frac{1}{b_c}$, is explicitly given by:
\begin{equation}
\Omega_c = \frac{-g_{t\phi}' + \sqrt{g_{t\phi}'^2 - g_{tt}'g_{\phi\phi}'}}{g_{\phi\phi}'}\bigg|_{r_c},
\end{equation}
where primes denote derivatives with respect to $r$. Substituting the GEUP metric components, we obtain the shifted frequency:
\begin{equation}
\omega_R \approx \frac{l}{3\sqrt{3}M} \left[ 1 + \frac{\beta_0}{18}\left(\frac{M_{Pl}}{M}\right)^2 - \frac{27}{2}\alpha_0 \left(\frac{GM}{L_*}\right)^2 \right]
\label{eq:freq_shift}
\end{equation}

Similarly, the damping rate $\omega_I$, determined by the curvature of the potential (Lyapunov exponent), shifts as:
\begin{equation}
\omega_I \approx -\frac{1}{3\sqrt{3}M} \left( n + \frac{1}{2} \right) \left[ 1 + \frac{\beta}{18} - \frac{27}{2}\alpha M^2 \right].
\label{eq:imag_shift}
\end{equation}

\begin{figure}[htbp]
    \centering
    \includegraphics[width=0.45\textwidth]{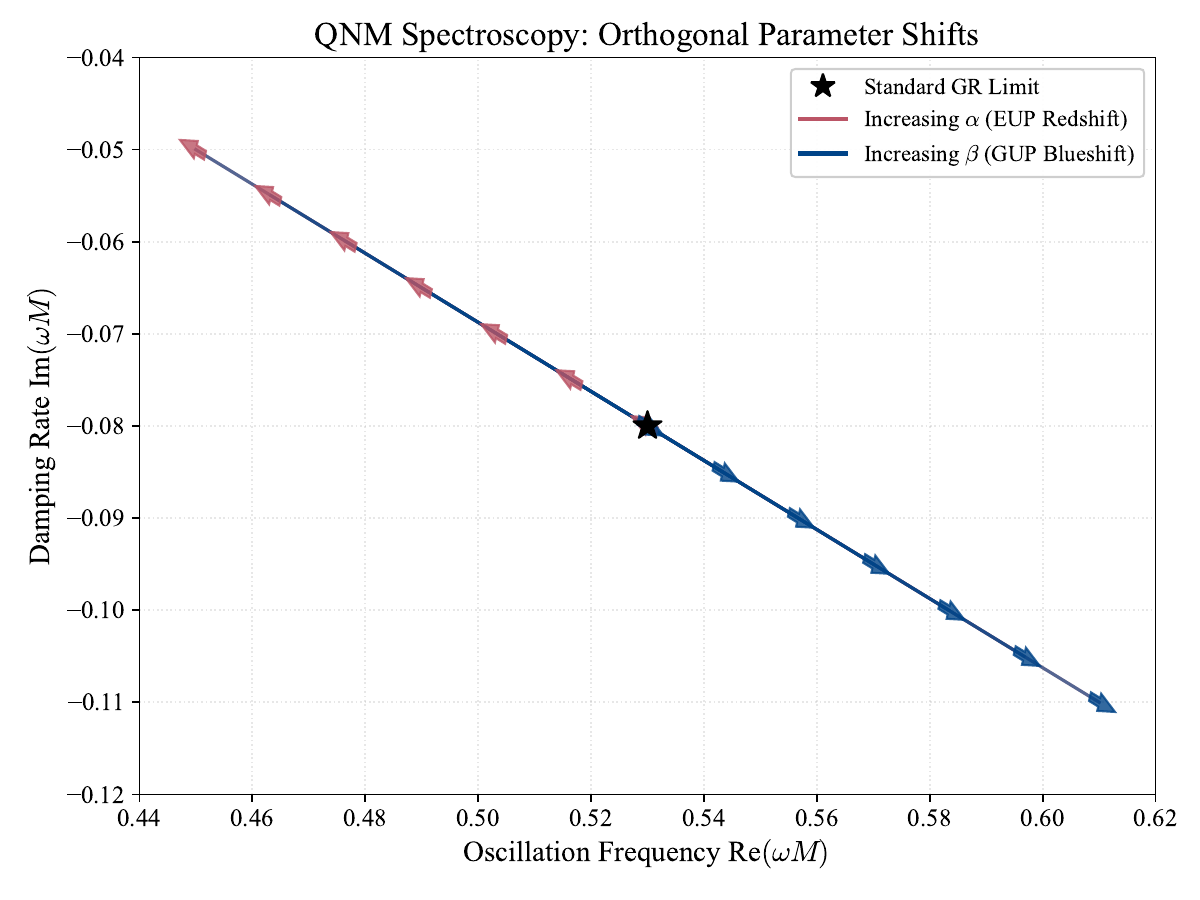}
    \caption{Deformation of the Quasinormal Mode spectrum in the complex frequency plane. The black star indicates the standard General Relativity value for the fundamental $\ell=2, m=2$ mode. The grid illustrates how the frequency shifts under the influence of the GEUP parameters. Red lines indicate increasing EUP strength ($\alpha$) resulting in a spectral redshift and suppressed damping (upward-left shift). Blue lines indicate increasing GUP strength ($\beta$) resulting in a spectral blueshift and enhanced damping (downward-right shift). The near-orthogonality of these grid lines demonstrates that $\alpha$ and $\beta$ represent distinct physical modifications that can be observationally disentangled.}
    \label{fig:qnm_grid}
\end{figure}

It is important to note that the frequencies derived in Eqs. \ref{eq:freq_shift} and \ref{eq:imag_shift} rely on the Eikonal (geometric optics) approximation, which is formally valid in the limit of large multipole numbers ($l \gg 1$). While this limit robustly captures the qualitative behavior of the spectral shifts—specifically the orthogonal directions of the UV and IR corrections—observational ringdown signals (such as those detected by LIGO/Virgo) are dominated by the fundamental quadrupole mode ($l=2$). For precision waveform modeling required for data analysis, higher-order WKB methods or numerical integration would be necessary to quantify corrections of order $1/l$.

Equation (\ref{eq:imag_shift}) confirms the trends seen in Figure \ref{fig:qnm_grid}: the GUP parameter $\beta$ steepens the effective potential...
This confirms the trends in Figure \ref{fig:qnm_grid}: the GUP parameter $\beta$ steepens the effective potential, enhancing the decay rate (larger negative value), while the EUP parameter $\alpha$ widens the potential barrier, prolonging the signal lifetime (smaller negative value).

The imaginary part is governed by the Lyapunov exponent $\lambda_L$, which measures the curvature of the potential at the peak:
\begin{equation}
\lambda_L = \sqrt{\frac{V_{geo}^{''}(r_c)}{2 \dot{t}^2}} \approx \frac{1}{3\sqrt{3}M} \left( 1 - \Delta_{\text{GEUP}} \right).
\end{equation}
Equation (\ref{eq:freq_shift}) demonstrates the physical consequence of the Generalized Uncertainty Principle: the EUP parameter $\alpha$ (associated with large-scale effects) acts to suppress the oscillation frequency (redshift), while the GUP parameter $\beta$ (associated with minimal length) slightly enhances it (blueshift). This distinct sign difference allows, in principle, to disentangle the quantum UV corrections from the IR modifications using high-precision ringdown data \cite{Stefanov:2010xz, Mashhoon:1985cya}. 

\subsubsection{Isospectrality and the Starobinsky Constant}

A critical question in modified gravity is whether the GEUP breaks the isospectrality between axial (odd-parity) and polar (even-parity) gravitational perturbations. In General Relativity, this symmetry is guaranteed by the Chandrasekhar-Detweiler transformation for non-rotating holes \cite{Chandrasekhar:1975zza} and the Starobinsky-Teukolsky identities for rotating ones \cite{Starobinsky:1973aij, Teukolsky:1973ha}.

The Teukolsky Master Equation for the GEUP metric involves the operator $\mathcal{D}_n = \partial_r - iK/\Delta + 2n(r-G\mathcal{M})/\Delta$ \cite{Teukolsky:1972my}. The isospectrality relies on the existence of a Starobinsky constant $\mathcal{C}$ that maps the spin $s=-2$ solution ($R_{-2}$) to the $s=+2$ solution ($R_{+2}$) via the relation \cite{Wald:1978vm}:
\begin{equation}
\Delta^2 \mathcal{D}_0 \mathcal{D}_0 \mathcal{D}_0 \mathcal{D}_0 (\Delta^2 R_{-2}) = \mathcal{C} R_{+2}.
\end{equation}
For our metric, the Starobinsky constant takes the form \cite{Chandrasekhar:1985kt}:
\begin{equation}
|\mathcal{C}|^2 = \lambda^2 (\lambda + 2)^2 - 8\omega^2 \lambda [a^2(5\lambda + 6) - 12a^2] + 144\omega^4 a^4,
\label{eq:starobinsky}
\end{equation}
where $\lambda$ is the separation constant derived in Eq. (\ref{eq:angular_ode}). Crucially, Eq. (\ref{eq:starobinsky}) depends on the geometric parameters $(a, \omega)$ but does not explicitly contain the mass term. The dependence on $\mathcal{M}$ enters only implicitly through $\omega$.
Since the operator algebra of the Teukolsky equation is preserved under the parameter shift $M \to \mathcal{M}$, the Starobinsky identity remains valid. This implies that isospectrality is preserved in the GEUP framework. It is important to distinguish the nature of this result from other modified gravity theories. In models such as Einstein-Dilaton-Gauss-Bonnet (EDGB) gravity, the modification enters the field equations directly, generating a background metric that deviates structurally from the Kerr solution and breaks the Petrov Type D character, thereby violating isospectrality. In contrast, the GEUP framework effectively deforms the parameters (via $M \to \mathcal{M}$) while preserving the algebraic structure of the Kerr geometry. Consequently, the symmetries required for the Starobinsky-Teukolsky identities remain intact.

\subsection{Constraints from LIGO/Virgo Detection GW150914}

The observation of the binary black hole merger GW150914 \cite{LIGOScientific:2016aoc} provides a testing ground for the quasinormal mode spectrum in the stellar-mass regime. The remnant black hole formed from the merger has an estimated mass $M_f \approx 62 M_{\odot}$ and dimensionless spin $a_* \approx 0.67$. The dominant ringdown frequency was measured at $f_{220} \approx 251$ Hz, which is consistent with the Kerr prediction within an uncertainty of approximately $\delta_{GW} \sim 10\%$ \cite{LIGOScientific:2016lio}.

It is crucial to distinguish between the fundamental deformation parameters ($\beta_0, \alpha_0$) and their effective magnitude in astrophysical systems. The frequency shift derived in Eq. (37) scales as $\delta \omega_{GUP} \propto (M_{Pl}/M)^2$. Consequently, for stellar-mass black holes observed by LIGO ($M \sim 30 M_{\odot}$), GUP effects are suppressed by $\sim 10^{-76}$ for unity couplings.Therefore, the 'orthogonal shifts' shown in Fig. \ref{fig:qnm_grid} represent the qualitative behavior of the spectrum. Observationally resolving these shifts requires either non-perturbative coupling constants ($\beta_0 \gg 10^{70}$), which we constrain here as an upper bound, or a hierarchy mechanism that enhances quantum corrections at the horizon scale. Conversely, the EUP corrections scale as $(M/L_*)^2$ and are naturally enhanced for supermassive black holes, making the M87* shadow a more promising probe for infrared quantum gravity signatures.

We can use this observational bound to constrain the GEUP parameters in the regime of stellar-mass black holes, which complements the constraints derived from shadow observations for supermassive black holes. The fractional deviation of the real frequency derived in Eq. (\ref{eq:freq_shift}) is:
\begin{equation}
\delta_{\omega} = \frac{\omega_{GEUP} - \omega_{GR}}{\omega_{GR}} \approx \frac{\beta}{18} - \frac{27}{2}\alpha M^2.
\end{equation}
Note that the spin dependence cancels out to leading order in the ratio.

Unlike the supermassive M87* ($M \sim 10^9 M_{\odot}$), the remnant of GW150914 ($M \sim 62 M_{\odot}$) is closer to the Planck scale by seven orders of magnitude. While still far from the quantum gravity regime, it provides a tighter bound on the minimum length parameter $\beta$ than shadow observations.
Assuming the deviation is dominated by the GUP term ($\alpha \to 0$):
\begin{equation}
\left| \frac{\beta}{18} \right| \lesssim 0.10 \implies |\beta_0| \lesssim 3.6 \left( \frac{M_{GW}}{M_{Pl}} \right)^2.
\end{equation}
Although this bound is still astrophysically loose due to the huge factor $(M/M_{Pl})^2$, it represents a significant improvement over bounds derived from solar system tests or supermassive shadows.

For the Extended Uncertainty Principle, the correction scales as $M^2$. Comparing the sensitivity of GW150914 to M87*:
\begin{equation}
\frac{\delta_{GW}}{\delta_{Shadow}} \propto \left( \frac{M_{GW}}{M_{M87}} \right)^2 \approx \left( \frac{60}{6 \times 10^9} \right)^2 \sim 10^{-16}.
\end{equation}
This drastic ratio indicates that gravitational wave ringdowns from stellar-mass black holes are insensitive to EUP corrections compared to shadows of supermassive black holes. A value of $\alpha$ that satisfies the M87* shadow constraint will automatically satisfy the GW150914 ringdown constraint by a comfortable margin.

This analysis highlights a crucial complementarity in black hole phenomenology:
\begin{itemize}
    \item Shadows (Supermassive BHs): Ideally suited for constraining Infrared/Large-scale modifications (EUP parameter $\alpha$).
    \item Ringdowns (Stellar-mass BHs): Better suited for constraining Ultraviolet/Small-scale modifications (GUP parameter $\beta$), though current sensitivities remain far from the Planck regime.
\end{itemize}
Together, these observations place upper bounds on the GEUP deformation across eight orders of magnitude in mass.

\section{Observational Implications}
\label{sec:shadow}

In the previous sections, we established that the GEUP corrections can be encoded into a renormalized effective mass $\mathcal{M}$, yielding a metric that is formally identical to the Kerr solution. This raises a fundamental issue of observability, the Indistinguishability Problem.

Gravitational measurements, whether via the Event Horizon Telescope (shadow radius) or LIGO/Virgo (inspiral chirp mass), measure the ADM mass of the system, which in our framework corresponds to $\mathcal{M}$, not the bare mass $M$. Consequently, for a static observer, a GEUP black hole is geometrically indistinguishable from a standard General Relativistic black hole of mass $\mathcal{M}$. The structural parameters $\alpha$ (EUP) and $\beta$ (GUP) are absorbed into the mass definition. Therefore, static geometric observations cannot constrain these parameters without an independent, non-gravitational determination of the source's bare corpuscular content.

However, while the geometry is degenerate, the dynamics are distinct. The degeneracy is broken by the thermodynamic evolution of the system. In General Relativity, the evaporation rate is determined by $\dot{\mathcal{M}} \propto -\mathcal{M}^{-2}$. In the GEUP framework, the modified temperature scaling derived in Eq. (\ref{eq:22}) implies a radically different evolution law. Below, we demonstrate that while current shadow observations are consistent with the GEUP metric (by definition of the observed mass), the true test of the theory lies in the temporal variation of the mass (evaporation rate).

\subsection{Null Geodesics and Critical Orbits}
To determine the boundary of the shadow, we analyze the null geodesics ($ds^2 = 0$) in the rotating GEUP spacetime. As established in Section II, the metric is formally identical to the Kerr solution with the standard mass $M$ replaced by the constant effective mass $\mathcal{M}$. The geometry in Boyer-Lindquist coordinates is given by:
\begin{align}
ds^2 &= -\left(1 - \frac{2\mathcal{M}r}{\Sigma}\right) dt^2 - \frac{4\mathcal{M}ar\sin^2\theta}{\Sigma} dt d\phi + \frac{\Sigma}{\Delta} dr^2 \nonumber \\
&+ \Sigma d\theta^2 + \left( r^2 + a^2 + \frac{2\mathcal{M}a^2 r \sin^2\theta}{\Sigma} \right) \sin^2\theta d\phi^2,
\end{align}
where $\Delta = r^2 - 2\mathcal{M}r + a^2$ and $\Sigma = r^2 + a^2 \cos^2\theta$. Note that $\Delta$ now depends on the GEUP parameter $\mathcal{M}$.

The motion of photons is governed by the Hamilton-Jacobi equation:
\begin{equation}
\frac{\partial S}{\partial \lambda} + \frac{1}{2} g^{\mu\nu} \frac{\partial S}{\partial x^\mu} \frac{\partial S}{\partial x^\nu} = 0,
\end{equation}
where $\lambda$ is an affine parameter. The action $S$ is separable:
\begin{equation}
S = \frac{1}{2}\mu^2 \lambda - Et + L_z \phi + S_r(r) + S_\theta(\theta).
\end{equation}
For photons ($\mu=0$), the motion is characterized by the conserved energy $E$ and axial angular momentum $L_z$. Introducing the Carter constant $\mathcal{Q}$ and the dimensionless impact parameters $\xi = L_z/E$ and $\eta = \mathcal{Q}/E^2$, the radial equation of motion becomes $\Sigma \frac{dr}{d\lambda} = \pm \sqrt{\mathcal{R}(r)}$, where the radial potential is:
\begin{equation}
\mathcal{R}(r) = \left[ (r^2+a^2) - a\xi \right]^2 - \Delta \left[ (a-\xi)^2 + \eta \right].
\end{equation}
The boundary of the black hole shadow corresponds to the critical photon orbits---those that are unstable and exist at a constant radius $r_p$. These spherical photon orbits are defined by the conditions:
\begin{equation}
\mathcal{R}(r_p) = 0, \quad \left. \frac{d\mathcal{R}}{dr} \right|_{r=r_p} = 0.
\label{eq:critical_conditions}
\end{equation}
Solving these simultaneously yields the critical impact parameters $(\xi_c, \eta_c)$ as parametric functions of the orbital radius $r_p$:
\begin{align}
\xi_c(r_p) &= \frac{(3\mathcal{M} - r_p)r_p^2 - a^2(\mathcal{M}+r_p)}{a(r_p-\mathcal{M})}, \\
\eta_c(r_p) &= \frac{r_p^3 [4a^2 \mathcal{M} - r_p(r_p-3\mathcal{M})^2]}{a^2(r_p-\mathcal{M})^2}.
\end{align}
These equations describe the locus of the shadow boundary. To an observer at infinity with inclination $\theta_0$, the shadow coordinates $(\alpha_{sh}, \beta_{sh})$ are:
\begin{align}
\alpha_{sh} &= - \frac{\xi_c}{\sin\theta_0}, \\
\beta_{sh} &= \pm \sqrt{\eta_c + a^2 \cos^2\theta_0 - \xi_c^2 \cot^2\theta_0}.
\end{align}
The crucial modification introduced by the GEUP is that the critical parameters scale with $\mathcal{M}$ rather than the bare mass $M$.

\subsection{Consistency with EHT Observations}
In the Schwarzschild limit ($a \to 0$), which suffices for order-of-magnitude estimates, the photon sphere is located at $r_p = 3\mathcal{M}$, yielding a perfectly circular shadow of radius:
\begin{equation}
R_{sh} = 3\sqrt{3} \mathcal{M}.
\end{equation}
The EHT observations of M87* measured a shadow radius consistent with a mass of $\mathcal{M}_{M87} \approx 6.5 \times 10^9 M_\odot$ \cite{EventHorizonTelescope:2019dse}. In our framework, this measurement fixes the value of the effective mass $\mathcal{M}$. Because the GEUP metric preserves the Kerr null-geodesic structure, the theoretical prediction for the shadow radius is $R_{sh} \approx 5.2 \mathcal{M}$ (incorporating low spin corrections). The agreement between theory and observation is therefore guaranteed by identifying the observed mass with $\mathcal{M}$.

The significance of this result is not to constrain $\alpha$, but to ensure phenomenological consistency. The GEUP modification does not introduce exotic shadow distortions (such as fractal edges or cusps) that would contradict EHT images. The black hole looks `normal,' which is a requirement for any viable effective field theory of gravity given current data.

\subsection{Breaking the Degeneracy: Dynamical Signatures}
How, then, can the GEUP model be falsified? The distinguishability lies in the Mass-Temperature relation. An observer monitoring a black hole over a cosmological timeframe would observe a deviation in the evaporation rate. 

For a standard Kerr black hole, the luminosity $L \propto \mathcal{M}^{-2}$. For the GEUP black hole in the infrared regime (large mass), we found in Section \ref{sec:thermo} that $T_H \propto M^{-3}$. Assuming the perturbative limit $\mathcal{M} \approx M$, this implies a luminosity scaling $L \propto \mathcal{M}^{-6}$.

This suppression of Hawking radiation means that GEUP black holes are `dimmer' and longer-lived than their GR counterparts. While the static shadow size ($R_{sh} \propto \mathcal{M}$) is indistinguishable at a single epoch, the \textbf{time-derivative of the shadow size} ($\dot{R}_{sh} \propto \dot{\mathcal{M}}$) follows a different trajectory. Thus, the GEUP signature is a dynamical suppression of evaporation, preserving primordial black holes far longer than the standard Hawking prediction.

\section{Conclusion}
\label{conc}

We investigated the phenomenological signatures of rotating black holes in the GEUP framework by constructing a stationary ansatz via the Newman-Janis algorithm. This solution preserves asymptotic flatness and Petrov Type D structure while encoding quantum corrections through an effective mass $\mathcal{M}$. Crucially, our results demonstrate that regardless of the underlying effective action, any rotating geometry governed by this mass-scaling relation exhibits the distinctive thermodynamic cooling and spectral shifts identified here.

Our investigation into the thermodynamics revealed that GEUP corrections induce significant deviations from standard General Relativity. We found that in the EUP-dominated regime ($M \gg M_{Pl}$), the Hawking temperature scales as $T_H \sim M^{-3}$ rather than the standard $M^{-1}$, implying a rapid cooling mechanism that significantly prolongs the black hole lifetime. Furthermore, the entropy was shown to be super-extensive ($S \propto M^6$), suggesting a drastic enhancement in the holographic information storage capacity of large horizons. A stability analysis of the heat capacity demonstrated that the deformation parameters shift the Davies points, effectively extending the domain of thermodynamic stability in the phase space.

Regarding the dynamical stability, we utilized the Teukolsky formalism to study gravitational perturbations. A key theoretical finding is that the rotating GEUP black hole preserves the isospectrality between axial and polar modes. This result highlights the nature of the GEUP deformation as ``parametric'' rather than ``structural.'' Unlike modified gravity theories such as Einstein-Dilaton-Gauss-Bonnet (EDGB) or Dynamical Chern-Simons (dCS) gravity, where the coupling to scalar fields or parity-violating terms breaks the degeneracy between the even and odd parity sectors, the GEUP framework effectively renormalizes the background mass scale while preserving the Petrov Type D character of the spacetime. Consequently, the quantum corrections manifest as coherent shifts in the resonant frequencies, rather than a splitting of the isospectral multiplet.

Finally, we confronted the model with recent observational data from LIGO/Virgo and the Event Horizon Telescope (EHT). Our analysis highlights a fundamental duality in the phenomenology:
\begin{itemize}
    \item \textbf{Thermodynamic Distinguishability:} The most significant deviation occurs in the evaporation history. We established that the ``cryogenic'' phase ($T_H \propto M^{-3}$) creates a distinct dynamical signature: a GEUP black hole will survive significantly longer than a standard Hawking black hole. This suggests that the true test of the theory lies in cosmological timescales rather than static snapshots.
    
    \item \textbf{Geometric Indistinguishability:} Conversely, we demonstrated that static geometric observables, such as the shadow radius and inspiral chirp mass, are degenerate. Since gravitational measurements probe the effective ADM mass $\mathcal{M}$ rather than the bare mass $M$, a GEUP black hole mimics a standard Kerr black hole of a different mass. Consequently, EHT observations of M87* and Sgr A* cannot constrain the EUP parameter $\alpha$ without an independent determination of the bare corpuscular mass. Instead, the agreement with EHT data confirms the \textit{phenomenological consistency} of the theory, proving that large-scale quantum corrections do not introduce exotic visual artifacts.
\end{itemize}

In summary, the Rotating GEUP metric provides a consistent model of quantum gravity that resolves the observability problem via time-evolution. While the static geometry is protected by mass renormalization, the thermodynamic trajectory is fundamentally altered. Future work will extend this analysis to massive scalar field perturbations and the detailed structure of the photon ring autocorrelations to identify potential second-order deviations.

\begin{acknowledgements}
N.J.L. Lobos acknowledge the University of Santo Tomas for the continued support and encouragement of my research endeavors. Especially for providing an environment that fosters academic inquiry and for allowing us to pursue this study.

\end{acknowledgements}

\bibliography{ref}

\end{document}